# Real-Time Stress Detection via Photoplethysmogram Signals: Implementation of a Combined Continuous Wavelet Transform and Convolutional Neural Network on Resource-Constrained Microcontrollers

Yasin Hasanpoor
*Advanced Service Robots (ASR) Lab., Department of Mechatronics Engineering,*
Faculty of New Sciences and Technologies, University of Tehran, Tehran, Tehran, Iran
yasin.hasanpoor@ut.ac.ir

Amin Rostami
*Advanced Service Robots (ASR) Lab., Department of Mechatronics Engineering,*
Faculty of New Sciences and Technologies, University of Tehran, Tehran, Tehran, Iran
aminrostami@ut.ac.ir

Bahram Tarvirdizadeh
*Advanced Service Robots (ASR) Lab., Department of Mechatronics Engineering,*
Faculty of New Sciences and Technologies, University of Tehran, Tehran, Tehran, Iran
bahram@ut.ac.ir

Khalil Alipour
*Advanced Service Robots (ASR) Lab., Department of Mechatronics Engineering,*
Faculty of New Sciences and Technologies, University of Tehran, Tehran, Tehran, Iran
k.alipour @ut.ac.ir

Mohammad Ghamari
*Department of Electrical and Computer Engineering,*
Kettering University,
Flint, Michigan, USA,
mghamari@kettering.edu

*Abstract*—This paper introduces a robust stress detection system utilizing a Convolutional Neural Network (CNN) designed for the analysis of Photoplethysmogram (PPG) signals. Employing the WESAD dataset, we applied Continuous Wavelet Transform (CWT) to extract informative features from wrist PPG signals, demonstrating enhanced stress detection and learning compared to conventional techniques. Notably, the CNN achieved an impressive accuracy of 93.7% after five epochs, post-implementation on a resource-constrained microcontroller. The optimization process, including pruning and Post-Train Quantization, was crucial to reduce the model size to 1.6 megabytes, overcoming the microcontroller's limited resources of 2 megabytes of Flash memory and 512 kilobytes of RAM. This optimized model not only addresses resource constraints but also outperforms traditional signal processing methods, positioning it as a promising solution for real-time stress monitoring on wearable devices.

## I. INTRODUCTION

The field of stress detection has witnessed substantial progress throughout its history, driven by an increasing understanding of the profound impact of stress on human health and well-being[1]. Early research focused on physiological measures such as heart rate and skin conductance[2]. However, the advent of wearable technology marked a transformative phase, prompting researchers to explore unobtrusive systems for continuous stress monitoring[3]. Efforts in stress detection have evolved alongside advancements in wearable devices, reflecting a persistent pursuit of real-time and lightweight systems to address the challenges of modern lifestyles[3]. Figure 1 illustrates the influence of stress on PPG signals.

The landscape of stress detection and emotion recognition through physiological signals has undergone significant evolution [4]. Prior research has delved into diverse physiological signals, employing various algorithmic models to address the complexities of this domain[5][6]. The selection of physiological signals and the algorithms employed often distinguishes studies in this area [7][4]. For example, investigations have explored EEG signals [4][8], EDA signals [9], HRV signals [10], blood pressure signals [11], and HR signals [12], among others. These studies collectively contribute to the comprehensive understanding of stress assessment using physiological data. In the context of the current research, emphasis is placed on PPG signals as a foundational element for stress detection. This choice aligns with previous works that have investigated the potential of PPG signals in stress assessment [13][14]. Figure 1 illustrates the impact of stress on PPG signals, elucidating why PPG signals are susceptible to stress and validating their efficacy as a reliable indicator for stress detection. The distinctive approach in our study integrates CWT analysis and CNNs to explore stress-related patterns within PPG data. This innovative combination expands the scope of stress assessment methodologies, holding promise for advancements in control, automation, and instrumentation within this domain.

The implementation of stress detection systems has been a focal point in the literature, representing a noteworthy breakthrough in stress research. CNNs, renowned for their prowess in recognizing patterns and extracting features from intricate physiological data [15][16], have emerged as a pioneering approach. Leveraging the capabilities of deep learning, this innovative methodology unveils intricate stress-related patterns within physiological signals [17]. The utilization of CNNs in stress detection represents a significant stride in enhancing the efficacy of stress assessment systems, providing valuable insights for the development of robust and accurate stress monitoring solutions [18][19].

The primary objective of this research is to address the pressing need for efficient stress detection in light and small-scale systems, particularly those with limited resources such as microcontrollers. The importance of such systems is underscored by their potential applications in daily life scenarios, driving safety, and health management, particularly for individuals prone to cardiovascular diseases. This paper contributes to the field by presenting a novel approach that combines CWT of PPG signals with a specifically designed CNN. The method is not only effective in stress classification but also optimized for implementation on resource-constrained microcontrollers, offering a solution that aligns with the demands of wearable devices.

Our methodology involves a multi-step process. Initially, we extract informative features from wrist PPG signals using CWT. Subsequently, a CNN is designed for stress classification. The innovative aspect of this approach lies in its adaptability to microcontroller environments, where Flash memory and RAM constraints necessitate careful optimization. This paper outlines the steps taken to implement and optimize the model for a microcontroller, ensuring its compatibility with lightweight systems. The methodology is validated through testing, demonstrating the effectiveness of the proposed stress detection system in real-world scenarios.

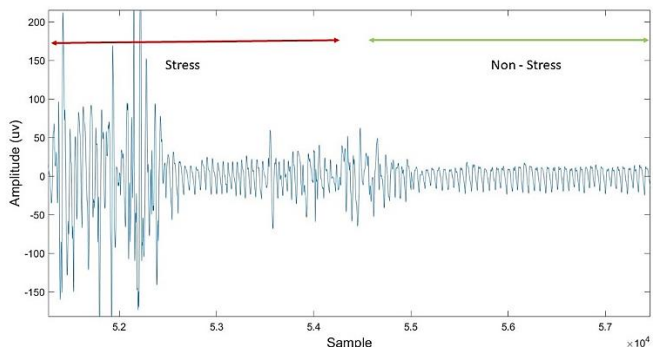

Figure 1. Effect of Stress on PPG Signals

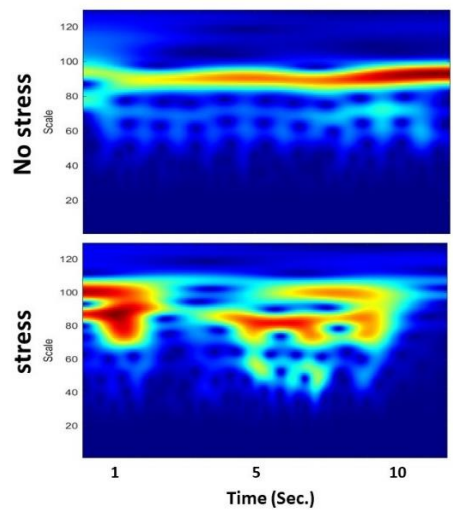

Figure 2. Comparative CWT Analysis of PPG Signals for Stress and Non-Stress Classes

## II. METHODOLOGY

The methodology section outlines the step-by-step process of constructing, training, optimizing, and evaluating the deep learning model for stress detection using PPG signals.

### A. Data Acquisition and Preprocessing

PPG signals are directly recorded from the sensor, focusing on wrist PPG signals with a sampling rate of 64 Hz. The raw PPG data is then preprocessed by segmenting it into 10-second windows with a 1-second stride, ensuring efficient feature extraction. Comparative CWT analysis (refer to Figure 2) is performed to illustrate the distinctive patterns in PPG signals for stress and non-stress classes.

### B. Data Splitting and Normalization

The dataset is divided into two sets: training and testing. The PPG images are loaded and processed to create numpy arrays. Data augmentation techniques, including rotation, width and height shift, and horizontal flip, are applied using an ImageDataGenerator to enhance the model's robustness. The training dataset is further augmented in batches to improve generalization.

### C. Model Architecture

The CNN model is designed to capture intricate patterns in the PPG signals. The architecture consists of two Conv2D layers with increased filters for feature extraction, followed by max-pooling layers to reduce spatial dimensions. A Flatten layer prepares the data for fully connected layers, which include a dense layer with increased neurons. The output layer uses the softmax activation function to classify stress and non-stress classes.

### D. Model Compilation and Training

The model is compiled using the Adam optimizer and sparse categorical crossentropy loss function. The training process involves the use of an augmented training generator, enhancing the model's ability to generalize. The model is trained for an increased number of epochs (5 in this case) to capture complex patterns in the data.

### E. Evaluation and Optimization

The model's architecture and performance metrics are displayed, showcasing the total and trainable parameters. Training history charts, depicting accuracy and validation accuracy over epochs, provide insights into the model's learning progression. The model is then evaluated on the testing dataset, and precision-recall curves are plotted to visualize its performance for each class.

### F. Optimization for Microcontroller

The model, initially requiring 19.3 MB of flash memory storage, undergoes optimization to meet the constraints of the microcontroller (STM32H743iit6). Pruning and post-training quantization techniques are sequentially applied, significantly reducing the model's size to 1.6 MB and streamlining RAM requirements to 302 KB (refer to Figure 4), ensuring compatibility with the microcontroller's limitations.

### G. Model Deployment and Inference

The optimized model is successfully deployed on the microcontroller. The final accuracy of 93.7% achieved on the microcontroller aligns with the original model's accuracy on Colab. Inference on the microcontroller takes 12.29 seconds per input image, demonstrating the efficiency of the deployed model in real-time scenarios.

The detailed process of building and optimizing the deep learning model on Colab and deploying it on the microcontroller is depicted in Figure 3. This flow chart visually represents the sequential steps involved in achieving an optimized model for deployment on a resource-constrained microcontroller.

the microcontroller maintains efficiency, taking 12.29 seconds per input image. The model is seamlessly converted from h5 format to tflm format using TensorFlow Lite for Microcontrollers, ensuring compatibility with the microcontroller's architecture. This meticulous methodology ensures the successful deployment of an optimized stress detection model on a lightweight microcontroller with limited resources.

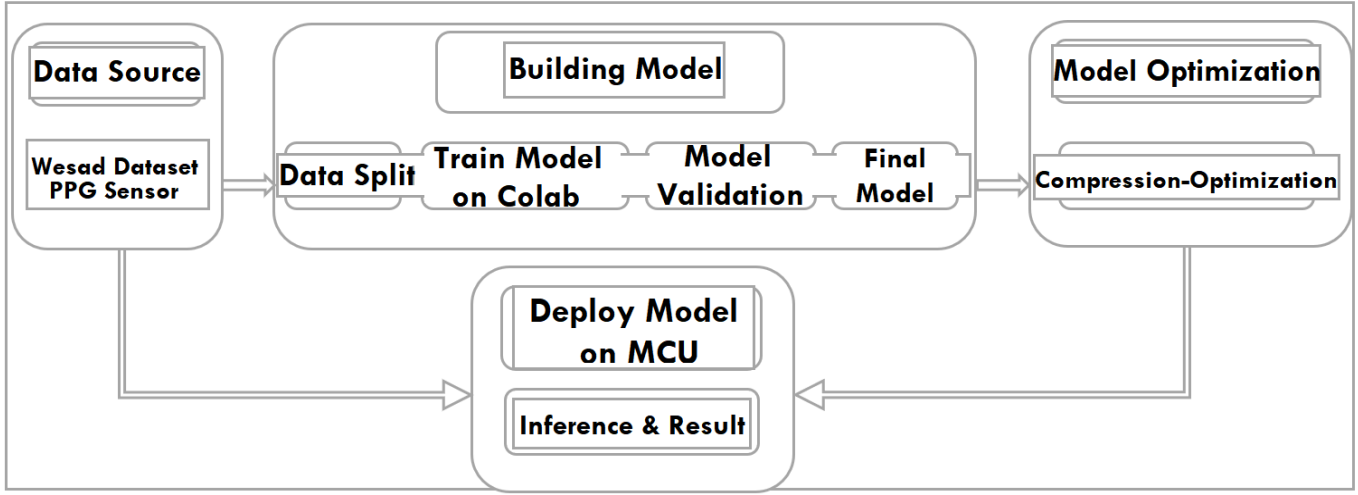

Figure 3. Flow chart of Building and Optimizing of our Deep Learning model on Colab and Deploying the built Model on MCU

### III. EXPERIMENTAL SETUP

The experimental setup section outlines the specifics of training and testing the optimized model, providing insights into the successful deployment on a microcontroller. The optimized model, requiring 19.3 MB of flash memory storage, initially exceeds the 2 MB limit of the STM32H743iit6 microcontroller. To address this limitation, a meticulous two-step optimization strategy is implemented. Firstly, pruning is applied to eliminate unnecessary weights and parameters, reducing the model size to one-third at 6.4 MB. Despite this reduction, the model size surpasses the microcontroller's flash memory capacity.

In the subsequent optimization step, post-training quantization is employed. This process significantly reduces the model size and computation RAM by converting model parameters and weights from 32-bit floats to 8-bit integers. Figure 4 illustrates the reduction in model size after both pruning and Post Quantization.

Post-optimization, the model size is reduced to one-fourth at 1.6 MB, allowing successful storage in the microcontroller's 2 MB flash memory. Moreover, RAM requirements for inference are streamlined, first to 3 MB after pruning, and eventually to 302 KB after quantization, well within the microcontroller's 512 KB RAM capacity.

The flow chart of figure 3 visually encapsulates the meticulous steps undertaken, emphasizing the seamless transition from model development to deployment on a microcontroller while ensuring optimal performance. This process not only highlights the adaptability of the model but also underscores the efficiency of the optimization techniques in preserving accuracy under constrained hardware conditions.

Figure 5 portrays the implementation and execution of the optimized model on the microcontroller. Despite resource constraints, the model demonstrates exceptional accuracy at 93.7%, identical to the original model on Colab. Inference on

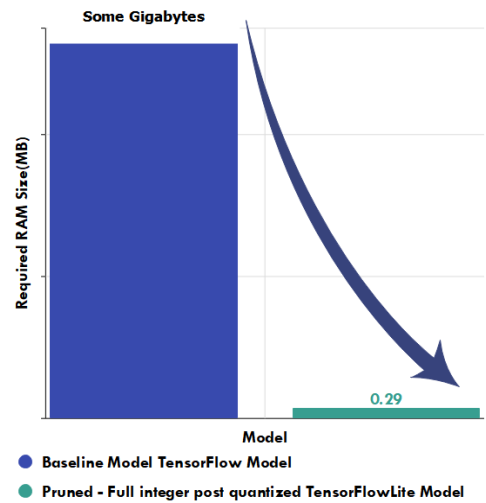

(a)

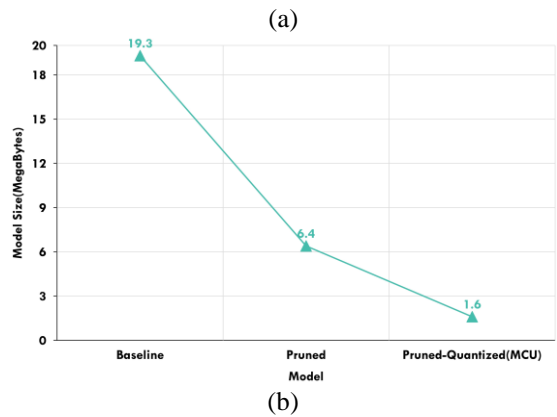

(b)

Figure 4. Comparison of RAM and Model Size (a) Required RAM size comparison between the baseline model and the pruned + post-quantized model. (b) Model size comparison at baseline, pruned, and quantized states.

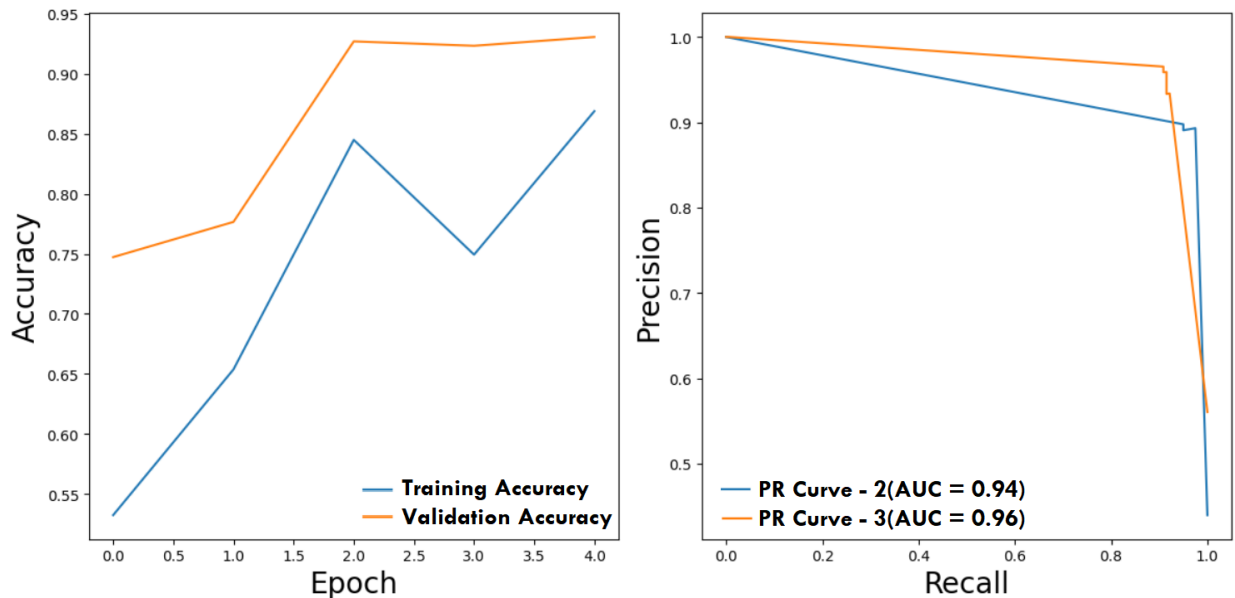

Figure 5. Training History and AUC Metrics

## IV. RESULTS

The results of the implemented deep learning model for stress detection using PPG signals are presented, showcasing the performance metrics, training history, and optimization achievements.

The optimized model achieved a remarkable validation accuracy of 93.7% on the microcontroller, maintaining parity with the original model trained on Colab. This robust performance demonstrates the effectiveness of the model in stress classification using wrist PPG signals.

The training history charts (refer to Figure 5) illustrate the progression of accuracy and validation accuracy over the training epochs. Notably, the model demonstrates rapid learning, with accuracy and validation accuracy surpassing 85% after the second epoch. This early convergence indicates the model's efficiency in learning stress-related patterns from the PPG signals.

The AUC values further validate the model's discriminative ability. The AUC of the training set is 0.94, emphasizing the model's proficiency in distinguishing stress and non-stress classes within the training data. Similarly, the AUC for the validation set is 0.97, reflecting the robustness and generalization capability of the model.

## V. CONCLUSION

In conclusion, this research presents a comprehensive approach to stress detection leveraging CWT on PPG signals and deploying a deep learning model optimized for a resource-constrained microcontroller.

The utilization of CWT provides valuable insights into the stress-induced variations in PPG signals, enabling the extraction of relevant features for effective stress classification. The designed CNN exhibits a high level of accuracy in stress detection.

The optimization process, including pruning and post-training quantization, successfully tailors the model to the constraints of a small microcontroller. The reduction in model size and RAM requirements ensures the feasibility of real-time stress detection in practical scenarios.

The training history, AUC values, and consistent accuracy across platforms underscore the model's reliability and generalization capabilities. The deployment on a microcontroller opens avenues for lightweight, real-time stress detection systems suitable for various applications, including daily life monitoring, driving safety, and health-related contexts.

The achievements of this research not only contribute to the field of stress detection but also showcase the potential for implementing efficient deep learning models on lightweight devices, paving the way for practical and unobtrusive stress monitoring solutions.